\documentclass[prb,twocolumn,showpacs,amsmath,amssymb,]{revtex4}

\usepackage{graphicx}
\usepackage{dcolumn}
\usepackage{bm}

\begin{document}

\title{Electronic properties of SnTe-class topological crystalline insulator materials\footnote{Project supported by the Ministry of Science and Technology of China (2016YFA0301000) and the National Natural Science Foundation of China (Grant No. 11334006).}}

\author{Wang Jianfeng$^{a)}$, Wang Na$^{a)}$, Huang Huaqing$^{a)}$, and Duan Wenhui$^{a)b)c)}$\footnote{Corresponding author. Email: dwh@phys.tsinghua.edu.cn}}
\address{${^{a)}}$Department of Physics and State Key Laboratory of Low-Dimensional Quantum Physics, Tsinghua University, Beijing 100084, People's Republic of China}
\address{$^{b)}$Institue for Advanced Study, Tsinghua University, Beijing 100084, People's Republic of China}
\address{$^{c)}$Collaborative Innovation Center of Quantum Matter, Tsinghua University, Beijing 100084, People's Republic of China}
\date{\today}

\begin{abstract}

The rise of topological insulators in recent years has broken new ground both in the conceptual cognition of condensed matter physics and the promising revolution of the electronic devices. It also stimulates the explorations of more topological states of matter. Topological crystalline insulator is a new topological phase, which combines the electronic topology and crystal symmetry together. In this article, we review the recent progress in the studies of SnTe-class topological crystalline insulator materials. Starting from the topological identifications in the aspects of the bulk topology, surface states calculations and experimental observations, we present the electronic properties of topological crystalline insulators under various perturbations, including native defect, chemical doping, strain, and thickness-dependent confinement effects, and then discuss their unique quantum transport properties, such as valley-selective filtering and helicity-resolved functionalities for Dirac fermions. The rich properties and high tunability make SnTe-class materials promising candidates for novel quantum devices.

\end{abstract}

\keywords{topological crystalline insulator, SnTe, surface states, mirror symmetry}

\pacs{73.20.At, 73.22.-f, 71.20.-b, 73.63.-b}

\maketitle

\draft

\vspace{2mm}

\section{Introduction}

In the past decade, a new field dubbed topological insulator (TI)\cite{hasan2010colloquium,qi2011topological,kane2005z,bernevig2006quantum} has caused enormous attentions in the condensed matter physics. Different from the earliest topological states in quantum Hall effect\cite{klitzing1980new,thouless1982quantized} which requires a strong magnetic field, TI is time-reversal-symmetry protected, and can be characterized by the $Z_2$ topological invariant\cite{kane2005z,moore2007topological,fu2007topological,fu2007topological2}. In TIs, the bulk insulating states are accompanied by metallic helical Dirac-like boundary states which are related to its bulk topology and protected by the time-reversal symmetry. With the novel properties of the edge states, such as high mobility, absence of backscattering and spin-polarized conductivity channels, TIs provide a promising platform for realizing new electronics and spintronics applications. The discovery of TIs also results in the realization of quantum anomalous Hall effect\cite{yu2010quantized,chang2013experimental}. Moreover, the proximity effect between TI and superconductor hosts non-Abelian Majorana fermion and opens a new venue for topological quantum computations\cite{fu2008superconducting}. On the materials front, all TI materials are narrow-gap semiconductors with inverted band gap at an odd number of time-reversal-invariant momenta (TRIMs)\cite{fu2007topological,fu2007topological2}. Two-dimensional (2D) TIs (also known as quantum spin Hall states) range from HgTe/CdTe\cite{bernevig2006quantum,konig2007quantum} and InAs/GaSb\cite{liu2008quantum,knez2011evidence} quantum well structures to the layered honeycomb lattice materials such as silicene\cite{liu2011quantum,liu2011low}, germanene\cite{liu2011quantum,liu2011low}, stanene\cite{xu2013large,liu2011low,tang2014stable}, their halogenides\cite{xu2013large,si2014functionalized}, and ultrathin Bi films\cite{liu2011stable,wada2011localized}. The 3D TIs materials include Bi$_{1-x}$Sb$_x$ alloys\cite{fu2007topological2,hsieh2008topological}, Bi$_2$Se$_3$-class materials\cite{zhang2009topological,chen2009experimental,xia2009observation}, half-Heusler compounds\cite{chadov2010tunable,lin2010half}, TlBiSe$_2$ family chalcogenides\cite{sato2010direct,kuroda2010experimental,chen2010single}, strained HgTe\cite{fu2007topological2,brune2011quantum}, $\alpha$-Sn\cite{fu2007topological2,barfuss2013elemental}, and bismuth-based III-V semiconductors\cite{huang2014nontrivial}, etc.\cite{autes2015novel,huang2016topological,bansil2016colloquium} Additional ways could convert the normal insulators into TIs, such as external strain\cite{liu2014manipulating}, chemical doping\cite{shi2015converting}.

The discovery of TIs stimulates the identification and searching for more topological quantum states. In TIs, the two branches of helical edge/surface states are related to each other by time-reversal symmetry, and they are degenerate at the TRIMs due to the Kramers' theorem thus forming the Dirac points. In principle, degeneracies can also come from other types of symmetries, such as particle-hole symmetry, crystalline symmetry, etc. Therefore, finding new topological phases protected by other symmetries is a new hot topic in this rapidly developing field\cite{chiu2015classification}. In 2011, Fu\cite{fu2011topological} proposed that crystalline symmetry can protect new types of topological states, which are called topological crystalline insulators (TCIs). The first theoretically predicted and experimentally realized TCI materials are IV-VI semiconductors, with SnTe as a representative\cite{hsieh2012topological,tanaka2012experimental,dziawa2012topological,xu2012observation}. The symmetry responsible for their topological character is the mirror symmetry. Due to the simple crystal structure, SnTe-class TCI materials have caused extensive concerns. Later, some other TCI materials are theoretically proposed in transition metal oxides with a pyrochlore structure\cite{kargarian2013topological} and anti-perovskite materials\cite{hsieh2014topological}; both of these two classes of materials are also topologically protected by mirror symmetry. Additionally, crystalline symmetry has been extended from the point group to non-symmorphic space group and a new class of TCIs named topological non-symmorphic crystalline insulator is proposed\cite{liu2014topological}. Considering the richness and complexity of crystal operations, recently some researchers\cite{jadaun2013topological,slager2013space,dong2015classification} have made great efforts to classify the TCIs systematically. However, these proposals ask for the support of realistic materials, and to date the only TCI materials realized in experiments are the IV-VI semiconductors. In this article, we mainly review the recent research on the SnTe-class TCI materials, which may facilitate future researches to search for more TCI materials and particularly to explore their potential applications.

In the research field of topological materials, the theoretical calculations play an important role in the material predictions and provide a reliable guidance for the experimental observations\cite{bansil2016colloquium}. So in the following (Sections II and III), we will first describe the topological identification of the bulk and surface states properties of SnTe from a computational point of view, and then we introduce the experimental observations. Due to the crystalline symmetry protection and multiple Dirac surface states, TCIs have more tunable properties than TIs. In Sections IV to VI, we will demonstrate the electronic properties of TCIs under various perturbations, carrier types control or superconductivity by chemical doping, strain-tuned surface states and topological phase transition, and thickness-dependent confinement effects on 2D topological states. Some novel quantum transport properties, such as valley-selective filtering and helicity-resolved functionalities for Dirac fermions, are  also discussed in Sections V and VI. Finally, we give the conclusions and outlook in Section VII.

\section{Bulk topology of SnTe class of materials}

SnTe-class IV-VI semiconductors have a rocksalt structure [Fig. 1(a)], and its Brillouin zone (BZ) is shown in Fig. 1(b). These materials are narrow-gap semiconductors, with their valence-band maximum (VBM) and conduction-band minimum (CBM) at four equivalent $L$ points of BZ. In the vicinity of $L$ point, the band structures of PbTe and SnTe with $p$ orbital projections of cation and anion based on first-principles calculations are presented in Figs. 1(c) and (d), respectively. It is seen that PbTe has a normal band ordering, i.e. The VBM is primarily derived from Te atoms and the CBM from Pb atoms, suggesting that PbTe can be smoothly connected to the atomic limit. While SnTe has an inverted band structure: the VBM is dominated by the Sn atoms and the CBM by Te atoms. The band inversion of SnTe relative to PbTe indicates that the relative ordering of $L_{6}^{+}$ and $L_{6}^{-}$ states is switched, as shown in Figs. 1(c) and (d). Nonetheless, it should be noted that there are four $L$ points in the whole BZ, leading to that band inversion occurs at an even number of points. So neither SnTe nor PbTe in the rocksalt structure is a TI with non-trivial $Z_2$ topological invariant.

\begin{figure}[ht]
\centering
\includegraphics[width=0.47\textwidth]{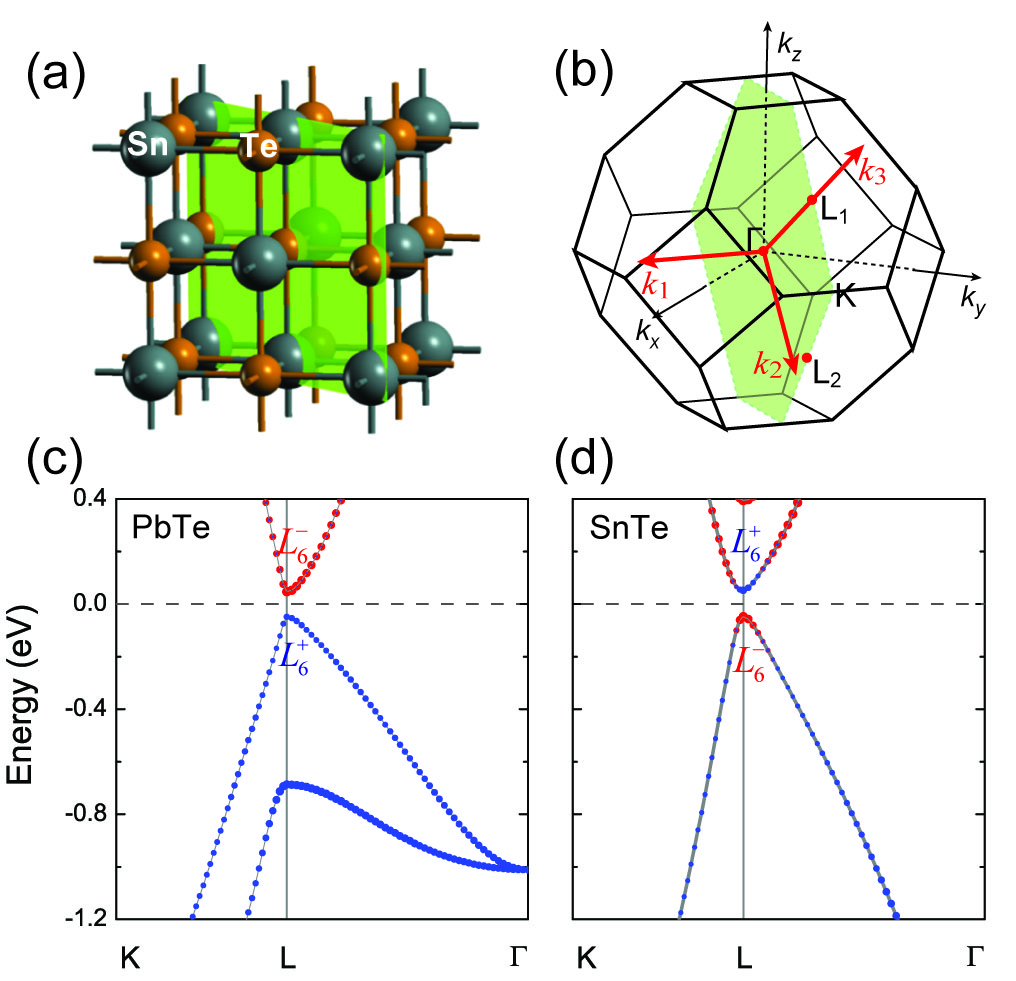}
\caption{\label{fig:fig1} (Color online) (a) The crystal structure of SnTe, in which the green plane depicts one of \{110\} mirror planes. (b)The face-centered-cubic Brillouin zone, in which the $\Gamma L_1L_2$ plane is invariant under reflection about the (110) mirror plane. (c) and (d), the band structures of PbTe and SnTe with $p$ orbital projections of cation (red dots) and anion (blue dots). The calculations are performed with optimal lattice constants (6.56 {\AA} for PbTe and 6.40 {\AA} for SnTe). At $L$, the lowest-lying valence and conduction states associated with even and odd parity eigenvalues are labeled by $L_{6}^{+}$ and $L_{6}^{-}$, respectively.}
\end{figure}

However, the band inversions of SnTe can give new non-trivial topology if we take the crystalline symmetry into account. Face-centered-cubic structure has the mirror symmetry with respect to the \{110\} planes, one of which is shown by the green plane in Fig. 1(a), corresponding to the plane $\Gamma L_1L_2$ in momentum space [Fig. 1(b)]. The Hamiltonian of this system under mirror operation $M$ satisfies $MH(k_1,k_2,k_3)M^{-1}=H(-k_1,k_2,k_3)$, in which $k_1$ is along the direction perpendicular to the plane $\Gamma L_1L_2$, and $k_3$ is along the direction of $\Gamma L_1$. On plane $\Gamma L_1L_2$, crystal momenta have $k_1=0$, and the Hamiltonian commutate with the mirror operator, i.e., $[H,M]=0$. Then these two operator have the common eigenstates. The mirror operation satisfies $M^2=-1$ for spin 1/2 electrons, and its eigenvalues are $M=\pm i$. So all the eigenstates of Hamiltonian can be labeled by the eigenvalues $\pm i$ of $M$, i.e., $\Psi_{+i}$ and $\Psi_{-i}$. The Hamiltonian can be written as the form of block diagonalization,
$H=\left(\begin{smallmatrix}
H_{+i} & 0 \\
0 & H_{-i} \\
\end{smallmatrix}\right)$.
Following the definition of spin Chern number\cite{sheng2006quantum}, here each class of eigenstates in subspace has an associated Chern number $C_{\pm i}$. Though the total Chern number is zero, $C=C_{+i}+C_{-i}=0$, due to the time-reversal symmetry, the mirror Chern number defined as $C_M=(C_{+i}-C_{-i})/2$ can be nonzero\cite{teo2008surface}. The crystal may have several mirror planes, and each can define an associated mirror Chern number. A nonzero mirror Chern number defines a topological crystalline insulator protected by the mirror symmetry\cite{hsieh2012topological}.

Hsieh {\it et al.}\cite{hsieh2012topological} introduced a low-energy effective model near the $L$ point to discuss the different topology between PbTe and SnTe, and they found that the band inversion of SnTe at $L$ changes the mirror Chern number by one. Because there are two $L$ points ($L_1$ and $L_2$) in the $\Gamma L_1L_2$ plane and these two $L$ points have the same contributions to the mirror Chern number\cite{hsieh2012topological}, finally the band inversion changes the mirror Chern number for the $\Gamma L_1L_2$ plane by two. As a result, PbTe is topologically trivial, while SnTe is a topological crystalline insulator with the mirror Chern number $C_M=-2$. Actually, there are six equivalent \{110\} mirror planes for SnTe, and each possesses two $L$ points, so the mirror Chern number for every plane is $-2$.

\section{Surface states of SnTe-class TCIs}

The bulk-boundary correspondence indicates the existence of topological surface states. Since band inversion occurs at four $L$ points in bulk BZ of SnTe, there should be four Dirac cones in the surface BZ. However, the gapless surface states of TCIs require the underlying crystalline symmetry to be preserved on the boundary. So not all crystal surfaces of SnTe have gapless surface states; only those surfaces that keep the mirror symmetry with respect to the (110) planes of rocksalt structure are gapless. Three common surface terminations are (001), (111), and (110)\cite{hsieh2012topological,ando2015topological,liu2013two}, and they meet the corresponding symmetry. Interestingly, there are two types of TCI surface states with qualitatively different electronic properties\cite{liu2013two}, which depend on the surface orientation and are schematically shown in Fig. 2. The type-I surface states have the properties that their Dirac points are located at TRIMs, such as the (111) surface states; while the Dirac points of type-II surface states on (001) and (110) surfaces are deviated from TRIMs. The latter types of surface states exhibit a Lifshitz transition as a function of Fermi energy. The dependence of boundary states on surface orientations is an important property of TCIs. Next we will first introduce the (001) surface states, and then the (111) surface states.

\begin{figure}[ht]
\begin{center}
\includegraphics[width=0.3\textwidth]{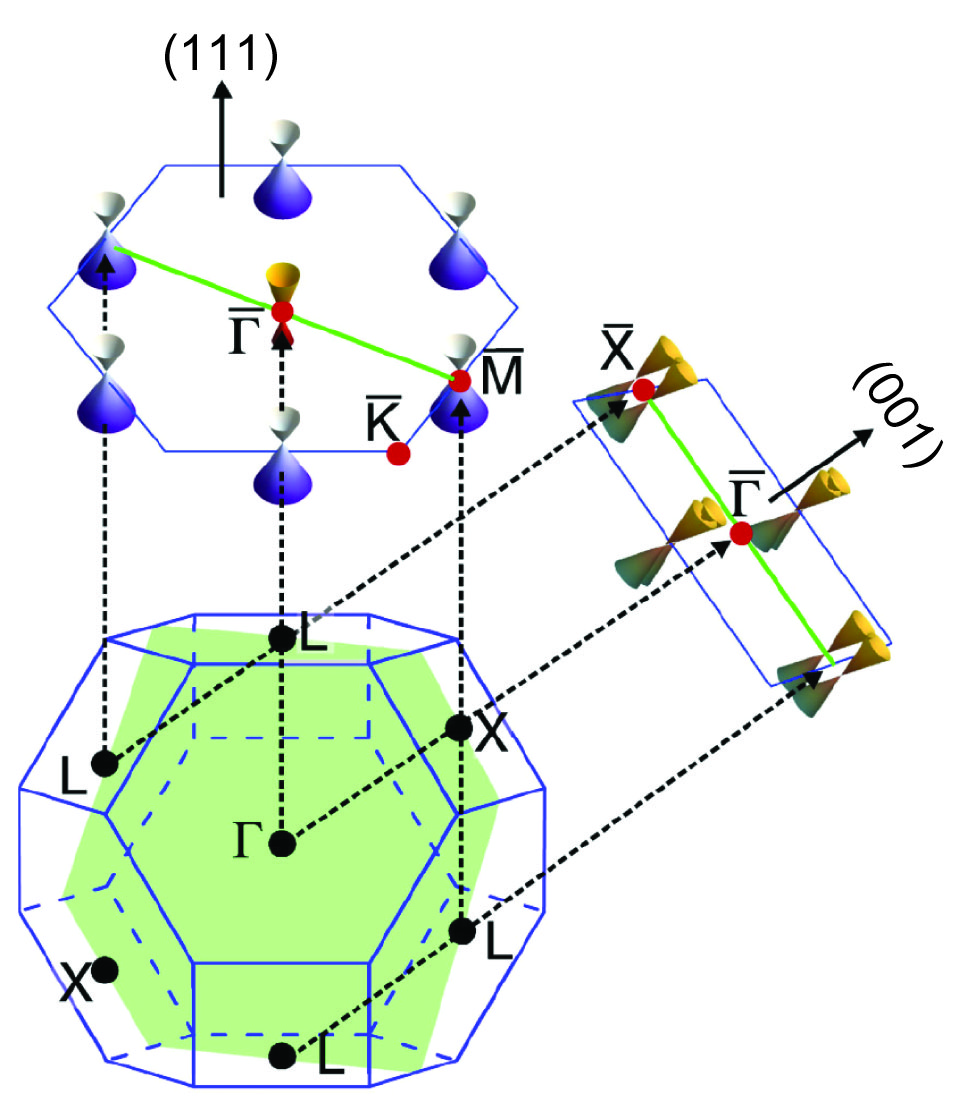}
\caption{\label{fig:fig2} (Color online) The projections of bulk BZ towards (001) and (111) surfaces of the rocksalt crystal structure, and the schematic locations of four Dirac cones in these two surface BZs.\cite{tanaka2013two}}
\end{center}
\end{figure}

\subsection{(001) surface states}

The first-principles calculated band structure of (001) surface of SnTe is shown in Fig. 3(a).\cite{hsieh2012topological} The red lines depict the topological surface band and they are of Dirac-like linear dispersion. Interestingly, the Dirac point is not located at the time-reversal invariant $\bar{X}$ point which is the projection of $L$ point in the bulk BZ, but has a little deviation along $\bar{\Gamma}\bar{X}$. As shown in Fig. 3(b), all four Dirac points in the whole surface BZ are deviated from $\bar{X}$, and they are related to each other by the $C_4$ rotation symmetry. Another important feature is that the change of Fermi surface topology exhibits a Lifshitz transition as a function of the Fermi energy. When the Fermi energy is slightly lower than the Dirac point, the Fermi surface consists of two disconnected hole pockets [red or orange elliptical region in Fig. 3(c)] on either side of the $\bar{X}$ point along $\bar{\Gamma}\bar{X}$; as the Fermi energy decreases, these two pockets enlarge and touch each other, then they reconnect to form a large hole pocket [green or blue region in Fig. 3(c)] and a small electron pocket [gray elliptical region in Fig. 3(c)], both centered at $\bar{X}$. In addition, this transition of Fermi surface is accompanied by a Van Hove singularity in the density of states arising from saddle points in the band structure.\cite{hsieh2012topological}

\begin{figure}[ht]
\begin{center}
\includegraphics[width=0.43\textwidth]{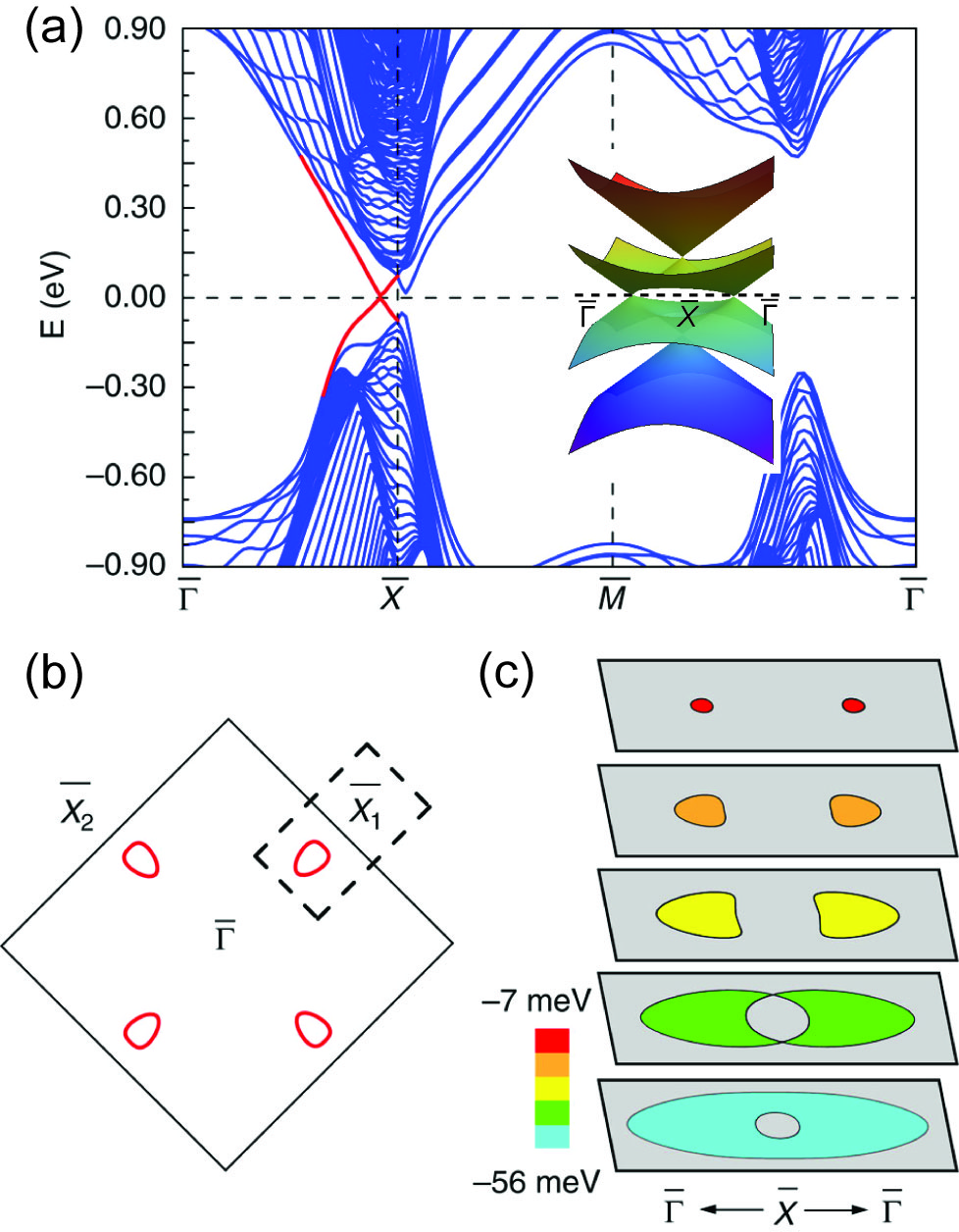}
\caption{\label{fig:fig3} (Color online) The (001) surface states of SnTe. (a) Band structure along high-symmetry lines from first-principles calculation. Inset: band structure of $k\cdot p$ model of Eq. (\ref{eq:H}) with 2D momentum near $\bar{X}$. (b) Fermi surface in the whole surface BZ. (c) A series of Fermi surfaces near the $\bar{X}$ point at different energies, displaying a Lifshitz transition.\cite{hsieh2012topological}}
\end{center}
\end{figure}

This type of TCI surface states can be understood as follows\cite{liu2013two,wang2013nontrivial}. From Fig. 2, we can see that two $L$ points ($L_1$ and $L_2$) in bulk BZ of SnTe are projected onto the same time-reversal invariant $\bar{X}$ point on the (001) surface. As a result of band inversions at these two $L$ points, two coexisting gapless Dirac cones will be created at $\bar{X}$. However, the sharp surface introduces intervalley scattering between these two $L$ points, thus producing the hybridization between the two massless surface Dirac fermions at the surface, and forming the novel surface states at $\bar{X}$. The essential properties of the (001) surface states can be captured by the following $k\cdot p$ Hamiltonian\cite{liu2013two,ando2015topological}:
\begin{equation}
H_{\bar{X}}(\textbf{k})=(v_xk_xs_y-v_yk_ys_x)\otimes I+m\tau_x+\delta s_x\tau_y.
\label{eq:H}
\end{equation}
Here, the first term describes two identical surface Dirac fermions associated with $L_1$ and $L_2$ (denoted by $\tau_z=\pm 1$); $\vec{s}$ is a set of Pauli matrices associated with the spin components; the second and third terms describe all possible intervalley hybridizations to zeroth order in $\textbf{k}$, which satisfy all the symmetries of the (001) surface.\cite{liu2013two,ando2015topological} The $m$ term induces the coupling between the two Dirac cones and produces an relative shift between them in energy; the $\delta$ term turns the band crossing of the two Dirac cones into an anticrossing via hybridization and opens a gap, which is, however, strictly forbidden along $\bar{\Gamma}\bar{X}$. Consequently, the bands are gapless along $\bar{\Gamma}\bar{X}$, but gapped along other directions. The band structure of $k\cdot p$ model of Eq. (\ref{eq:H}) with 2D momentum near $\bar{X}$ is shown as an inset in Fig. 3(a). The Hamiltonian above can perfectly repeat the first-principles electronic structure of (001) surface, such as the deviation of Dirac point from $\bar{X}$, the Lifshitz transition of Fermi surface, and the Van Hove singularity in the density of states. The Lifshitz transition is interesting, because it would accompany a marked change in the Dirac-carrier properties and provide another ingredient in the physics of topological materials.

\begin{figure*}[ht]
\begin{center}
\includegraphics[width=0.75\textwidth]{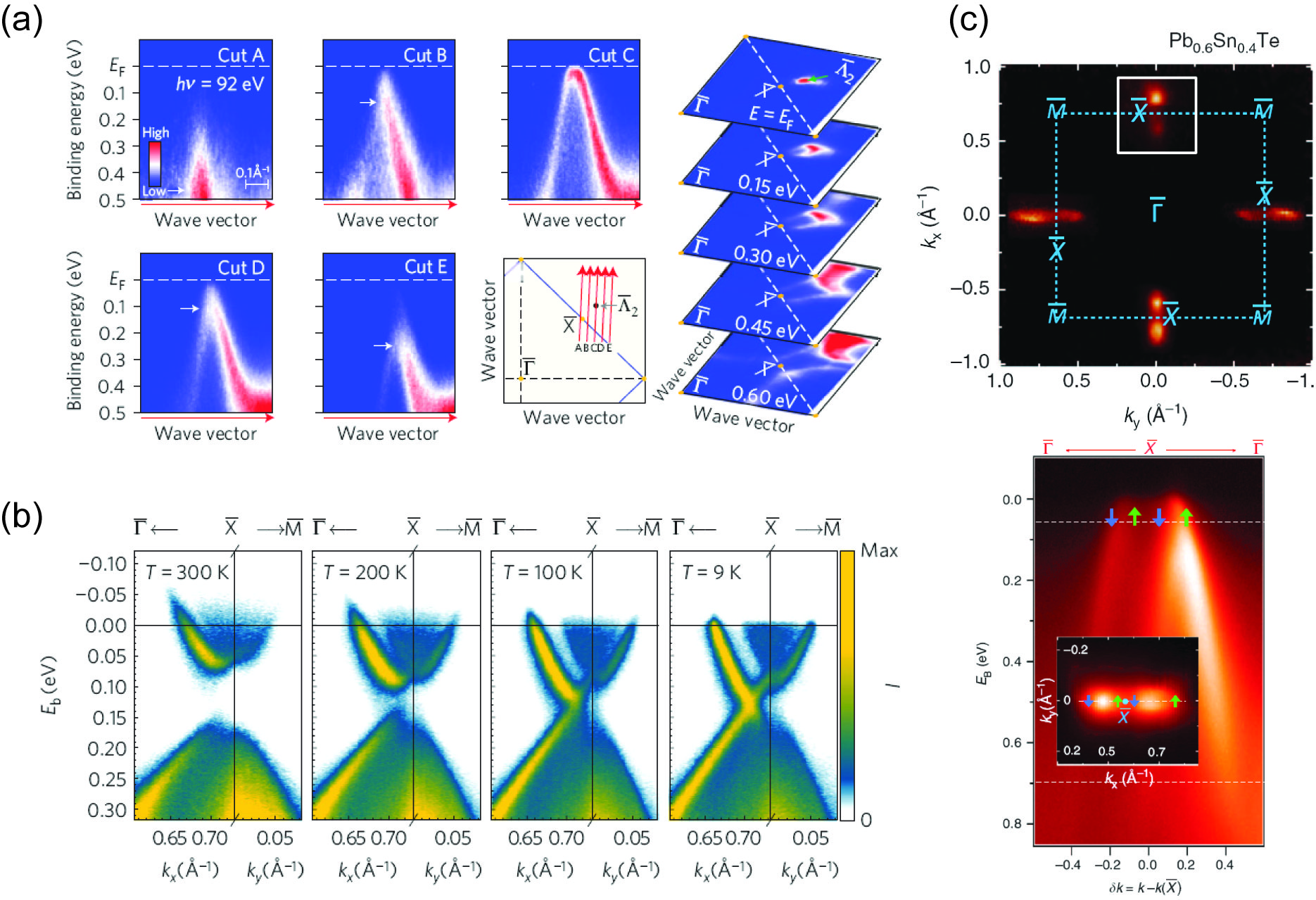}
\caption{\label{fig:fig4} (Color online) The ARPES measurements of the (001) surface states of SnTe-class materials. (a) ARPES intensity for SnTe along the cuts A--E in the surface BZ and mapping in a 2D wave-vector plane at various $E_{\rm B}$.\cite{tanaka2012experimental} (b) The temperature dependence of the ARPES data for the (001) surface of Pb$_{0.77}$Sn$_{0.23}$Se monocrystals. They clearly show the evolution of the gapped surface states (for $T\geq 100$ K) into the Dirac-like states on lowering the temperature ($T=9$ K).\cite{dziawa2012topological} (c) Up panel: ARPES iso-energetic contour mapping ($E_{\rm B}=0.02$ eV) of Pb$_{0.6}$Sn$_{0.4}$Te covering the first surface BZ. Down panel: ARPES dispersion map along the mirror line direction. Inset: Measured spin polarization profile is shown by the green and blue arrows on top of the ARPES iso-energetic contour at binding energy $E_{\rm B}=0.06$ eV.\cite{xu2012observation}}
\end{center}
\end{figure*}

As the cleavage plane, the (001) surface of SnTe and its family materials has been successfully prepared and the corresponding type-II surface states have been confirmed by experiments\cite{tanaka2012experimental,dziawa2012topological,xu2012observation}. Tanaka {\it et al.}\cite{tanaka2012experimental} grew the high-quality single crystal of SnTe, and observed the signature of a metallic Dirac-cone surface band by the angle-resolved photoemission spectrum (ARPES), with its Dirac point slightly away from the edge of the surface BZ [Fig. 4(a)]. In fact, because of the hole-doped nature of SnTe crystal, the Dirac point of surface states is not detected in their work and its energy is estimated to be 0.05 eV above $E_{\rm F}$ from a linear extrapolation of the two dispersion branches. In line with the theoretical prediction, they did not observe such gapless surface states in the cousin material PbTe. About the p-type conductivity of SnTe, we will explain its microscopic origin later in the article. For the Pb$_{1-x}$Sn$_x$Te and Pb$_{1-x}$Sn$_x$Se alloys, the chemical potential can easily be tuned to yield n-type or p-type conductivity by changing the crystals' growth condition, which makes them more suitable for experimental investigations on the TCI state. For example, Dziawa {\it et al.}\cite{dziawa2012topological} prepared the Pb$_{0.77}$Sn$_{0.23}$Se single crystal, and the ARPES of its (001) surface shows a n-type nature of this material. Importantly, they demonstrated that this alloy will undergo a temperature-driven topological phase transition from a trivial insulator to a TCI [Fig. 4(b)]. Xu {\it et al.}\cite{xu2012observation} utilized ARPES and spin-resolved ARPES to study the low-energy electronic structure below and above the band inversion topological transition of Pb$_{1-x}$Sn$_x$Te ($x=0.2$ and $x=0.4$). They demonstrated the observation of spin-polarized surface states in the inverted regime and their absence in the non-inverted regime, in which the ARPES iso-energetic contour mapping in the first surface BZ and spin-resolved topological Dirac surface states of Pb$_{0.6}$Sn$_{0.4}$Te are shown in Fig. 4(c). The interplay between topology and crystal symmetry is the character of TCIs. If the mirror symmetry of SnTe is broken, the gapless surface states will be gapped, i.e., the massless Dirac fermion will acquire mass. Okada {\it et al.}\cite{okada2013observation} reported high-resolution scanning tunneling microscopy studies of Pb$_{1-x}$Sn$_x$Se that reveal the coexistence of zero-mass Dirac fermions protected by crystal symmetry with massive Dirac fermions consistent with crystal symmetry breaking.

\subsection{(111) surface states}

Theoretically, the (111) surface states of SnTe are more simple than those of (001) surface. As shown in Fig. 2, four different $L$ points of bulk BZ are projected to different momenta in surface BZ, i.e., one $\bar{\Gamma}$ and three $\bar{M}$ points. There are no interactions between them. In fact, due to the polarity of (111) surface, it is not so easy to achieve the ideal (111) surface and observe its topological surface states in experiments. On the theoretical computations, there are only some tight-binding calculations at first without regard to the surface dangling bonds. Fig. 5 shows the surface band structure with two different terminations of SnTe (111) surface calculated by Liu {\it et al.}\cite{liu2013two} It is seen that Dirac-like surface band crossing (denoted by the red lines) occurs at $\bar{\Gamma}$ and $\bar{M}$ for both two terminations, whereas the Dirac points are close to the top (bottom) of the valence (conduction) band for Sn (Te) termination. The Dirac fermion at $\bar{\Gamma}$ is isotropic: the Fermi velocity along $\bar{\Gamma}\bar{M}$ is identical to that along $\bar{\Gamma}\bar{K}$; while at $\bar{M}$, the Dirac fermion is anisotropic: the Fermi velocity along $\bar{M}\bar{K}$ is larger than that along $\bar{M}\bar{\Gamma}$. The $k\cdot p$ Hamiltonians at $\bar{\Gamma}$ and $\bar{M}$ are given by $H_{\bar{\Gamma}}(\textbf{k})=v(k_1s_2-k_2s_1)$ and $H_{\bar{M}}(\textbf{k})=v_1k_1s_2-v_2k_2s_1$, where $k_1$ is along the $\bar{\Gamma}\bar{K}$ direction and $k_2$ is along the $\bar{\Gamma}\bar{M}$ direction.\cite{liu2013two}

\begin{figure}[ht]
\begin{center}
\includegraphics[width=0.47\textwidth]{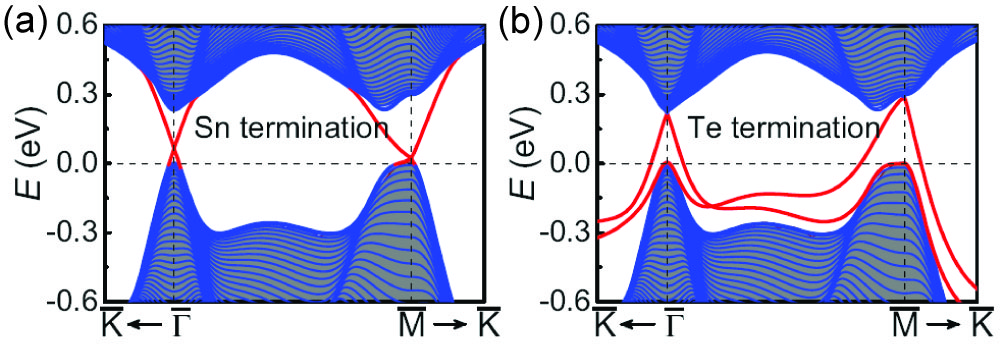}
\caption{\label{fig:fig5} (Color online) Band structures of the (111) surface for Sn and Te terminations. There are four Dirac pockets in the surface BZ: one at $\bar{\Gamma}$ and three at $\bar{M}$.\cite{liu2013two}}
\end{center}
\end{figure}

The ideal (111) surface of SnTe-class materials is a polar surface with unpaired surface electrons. In principle, the huge electrostatic potential induced by the dipole accumulation will cause the instability of the surface\cite{tasker1979stability}. Using first-principles calculations, Wang {\it et al.}\cite{wang2014structural} studied the (111) surface morphology and the associated electronic structures of SnTe under different growth conditions. They found that three different surfaces can stably exist, with the increasing of chemical potential of Sn: unreconstructed Te-terminated surface, $(\sqrt{3}\times\sqrt{3})$-reconstructed surface and $(2\times 1)$-reconstructed surface. The electronic structure of unreconstructed (111) surface shows that the nontrivial surface states are affected by the unpaired electrons, however clean type-I surface states can be achieved by hydrogen adsorption. Importantly, they revealed the critical effect of surface reconstruction on the topological surface states: the reconstruction will result in the folding of surface BZ and probably induce the breaking of the mirror symmetries, both of which can change the characteristics of topological surface states. For example, in Fig. 6(a,b), the $(2\times 1)$-reconstructed surface has the type-II surface states at $\bar{\Gamma}$: the Dirac point is deviated from TRIM. The $(2\times 1)$ reconstruction folds the surface BZ (SBZ) from hexagonal SBZ1 to rectangular SBZ2, causing the $\bar{M_1}$ point in SBZ1 to fold to the $\bar{\Gamma}$ point in SBZ2. As a result, two different $L$ points in the bulk BZ are projected to the same momentum $\bar{\Gamma}$ for reconstructed surface, just like the (001) surface. The additional intervalley scattering and the only one symmetric (110) mirror plane remaining create the type-II surface states at $\bar{\Gamma}$, which will also undergo a Lifshitz transition when tuning Fermi energy. Meanwhile the mirror symmetry breaking gaps the surface states at $\bar{M_2}$. In $(\sqrt{3}\times\sqrt{3})$-reconstructed surface, as shown in Fig. 6(c,d), the reconstruction only folds three different $\bar{M}$ in SBZ1 to three new $\bar{M^{\prime}}$ in SBZ3, so four Dirac cones still exist at four TRIMs belonging to type-I surface states. However, due to the different energy of Dirac points at $\bar{\Gamma}$ and $\bar{M}$, their conductive properties can be different when changing the Fermi energy.

\begin{figure}[ht]
\begin{center}
\includegraphics[width=0.48\textwidth]{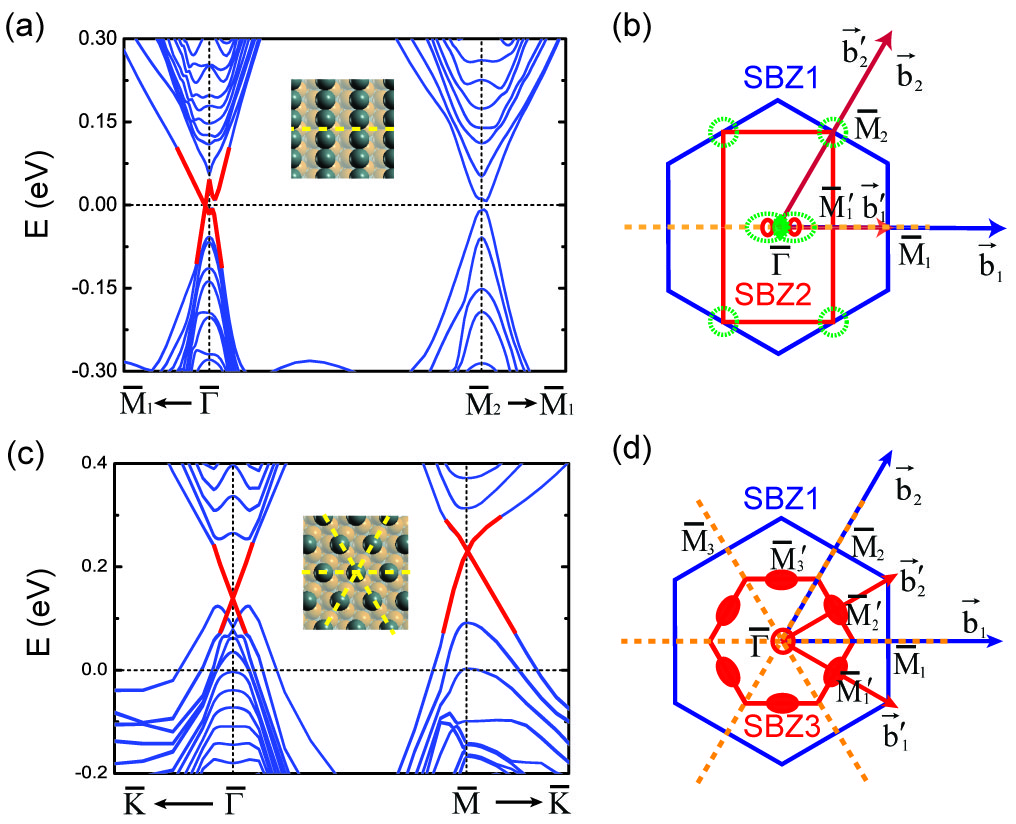}
\caption{\label{fig:fig6} (Color online) (a) Band structure of $(2\times 1)$-reconstructed (111) surface, where the surface states are denoted by red lines. Inset: surface atomic configuration, where only one of original three \{110\} mirror symmetries (indicated by yellow line) remains after reconstruction. (b) The surface Brillouin zones (SBZs): SBZ1 (blue hexagon) for original unreconstructed (111) surface and SBZ2 (red rectangle) for (2$\times$1)-reconstructed surface structure. The iso-energetic contour with energies at 5 meV (30 meV) above the Fermi energy is represented by red circle (green dotted circle and shaded area). (c) Band structure of $(\sqrt{3}\times\sqrt{3})$-reconstructed surface. Three mirror symmetry planes still exist (yellow lines in the inset). (d) SBZ3 and schematic iso-energetic contour with energy between the two Dirac points of $\bar{\Gamma}$ and $\rm{\bar{M}}$.\cite{wang2014structural}}
\end{center}
\end{figure}

In experiments, the observation of (111) surface states is truly more challenging owing to the difficulty in the preparation of non-cleavage surface. To date, epitaxial layer growth is an effective method for obtaining the (111) film. Fig. 7(a) and (b) show the ARPES measurements on the (111) surface of SnTe and Pb$_{1-x}$Sn$_x$Te performed by Tanaka {\it et al.}\cite{tanaka2013two} and Yan {\it et al.}\cite{yan2014experimental}, respectively. For the pure SnTe, both two experiments demonstrate the p-type conductivity, and the spectrum of surface states is covered by the signals of bulk valence band. Tanaka {\it et al.} utilized the second derivatives of the momentum distribution curves and changed the photon energies to recognize the linear-dispersion surface band. While Yan {\it et al.} changed the Pb/Sn ratio and found a topological phase transition in Pb$_{1-x}$Sn$_x$Te. A clear observation of the (111) surface states is done by Polley {\it et al.}\cite{polley2014observation} on the Pb$_{1-x}$Sn$_x$Se alloys. From Fig. 7(c), the Dirac-like band crossing is unchanged when varying the photo energy, revealing the character of surface band. They also observed that there is no relative binding energy difference between the $\bar{\Gamma}$ and $\bar{M}$ Dirac points. In addition, Wang {\it et al.}\cite{wang2015molecular} obtained the metastable rocksalt SnSe (111) films with the molecular beam epitaxy (MBE), and demonstrated that the epitaxial rocksalt SnSe is also a TCI through observing its (111) surface states.

\begin{figure}[ht]
\begin{center}
\includegraphics[width=0.46\textwidth]{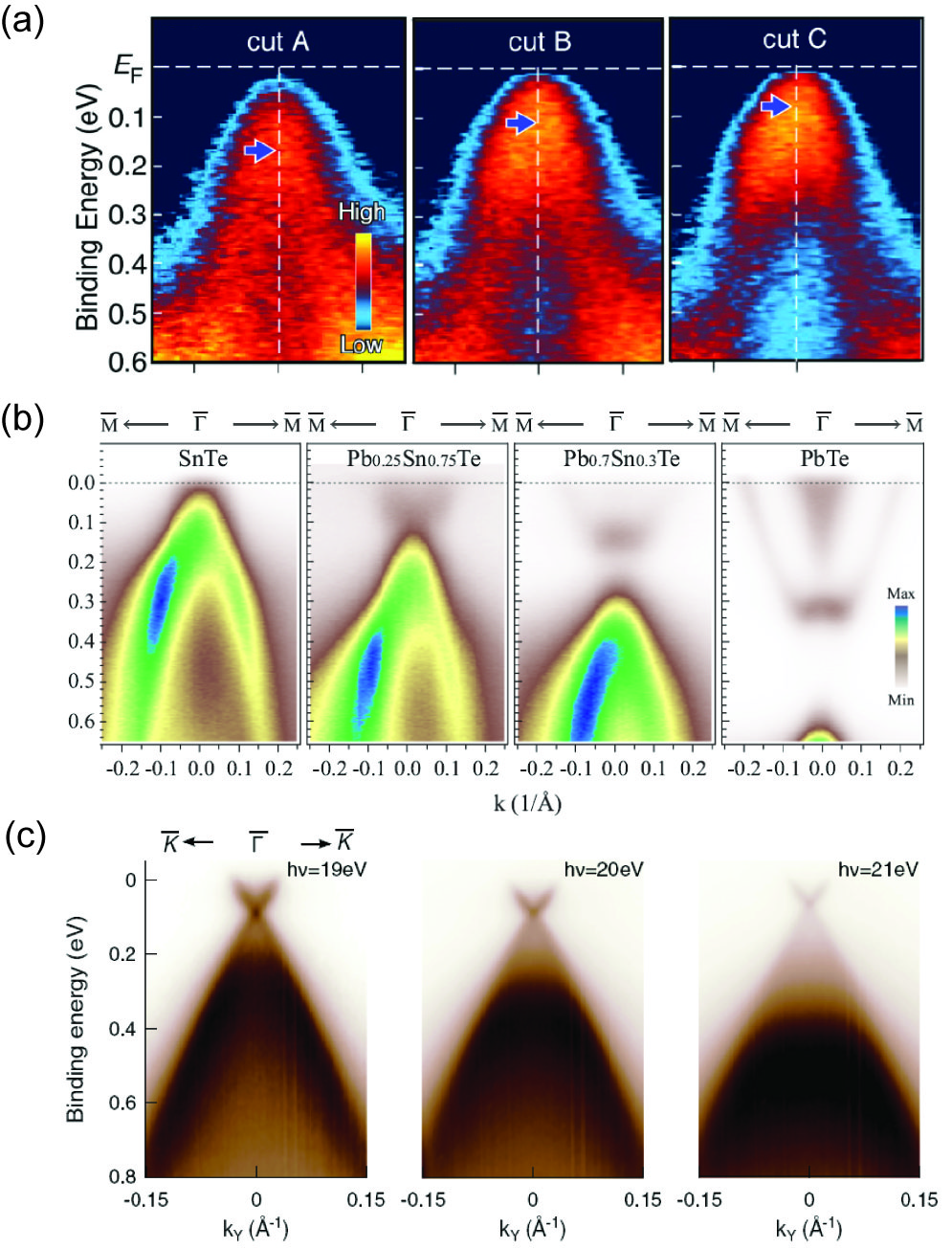}
\caption{\label{fig:fig7} (Color online) The ARPES measurements of the (111) surface states of SnTe-class materials. (a) Near-$E_{\rm F}$ ARPES intensity of SnTe measured along several cuts around the $\bar{\Gamma}$ point. Blue arrow indicates the top of the band.\cite{tanaka2013two} (b) ARPES intensity in the vicinity of the $\bar{\Gamma}$ point for SnTe, Pb$_{0.25}$Sn$_{0.75}$Te, Pb$_{0.7}$Sn$_{0.3}$Te, and PbTe, respectively.\cite{yan2014experimental} (c) Photon energy dependent ARPES measurements around the $\bar{\Gamma}$ point on a (111) oriented Pb$_{1-x}$Sn$_x$Se film. The bulk band is strongly dependent on the photon energy, while the position of the Dirac-like dispersion is unchanged when varying the photon energy, indicating the nature of surface states. However, its intensity is strongly modulated with the photon energy.\cite{polley2014observation}}
\end{center}
\end{figure}

\section{Native defects and doping in SnTe-class materials}

As mentioned above and observed by experiments, the as-grown SnTe crystals are usually heavily hole doped, which hinders the detection and investigations of the surface states and possible applications. Meanwhile, PbTe, the cousin material, can be either n or p type, depending on the growth conditions. Wang {\it et al.}\cite{wang2014microscopic} studied the role of native point defects in the doping behavior of these two materials and explained the microscopic origin of the p-type conductivity of SnTe.

As shown in Fig. 8(a), the dominant defect in SnTe is the negatively charged cation vacancy (V$_{\rm Sn}$$^{2-}$) under both Sn-rich and Te-rich conditions, and V$_{\rm Sn}$$^{2-}$ has a negative formation energy throughout the bulk gap. As a result, the Fermi level of SnTe has to be pushed under the VBM [i.e., a minus value of Fermi level in Fig. 8(a)] until the formation energy of V$_{\rm Sn}$$^{2-}$ becomes positive. So SnTe is always hole doped. While in PbTe [Fig. 8(b)], the dominant defect is not always the cation vacancy (V$_{\rm Pb}$$^{2-}$), but could be either anion or cation vacancy V$_{\rm Te}$$^{2+}$, V$_{\rm Pb}$$^{2-}$ or antisite Te$_{\rm Pb}$$^{2+}$, depending on the growth conditions. V$_{\rm Te}$$^{2+}$ and Te$_{\rm Pb}$$^{2+}$ are positively charged so that PbTe could be electron doped when these defects dominate.

\begin{figure}[ht]
\begin{center}
\includegraphics[width=0.47\textwidth]{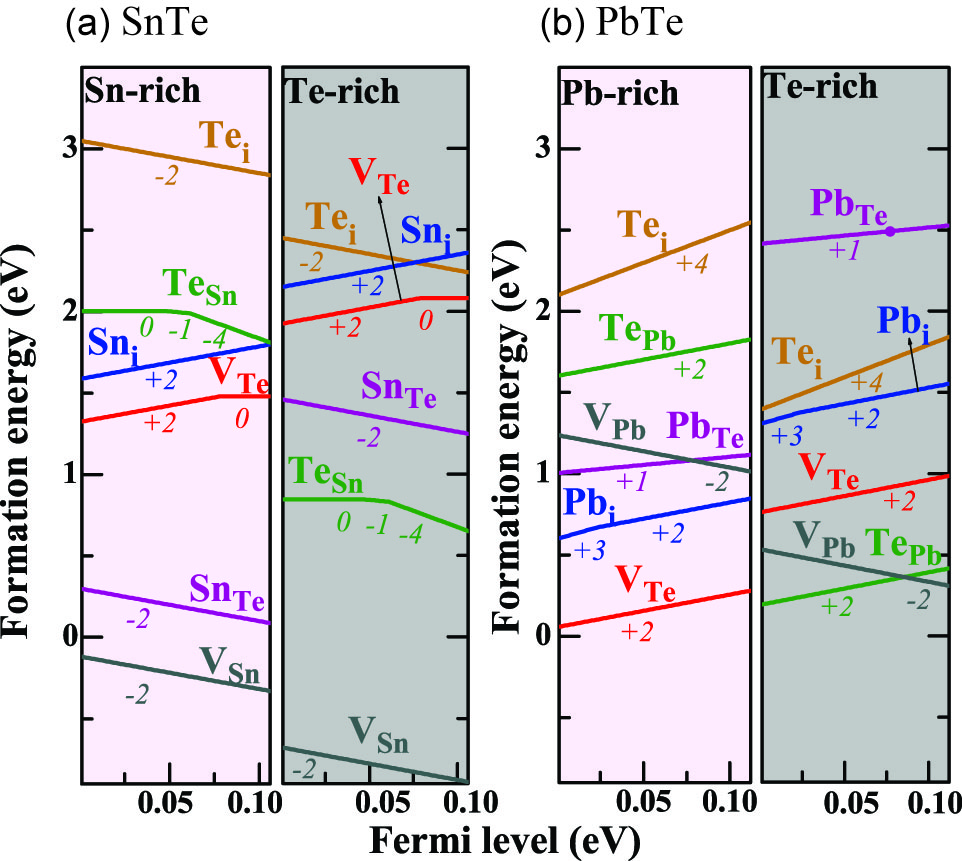}
\caption{\label{fig:fig8} (Color online) Formation energy as a function of Fermi level for native point defects in (a) SnTe and (b) PbTe under Sn/Pb rich and Te-rich conditions. The slope of each segment indicates the charge on the defect. Zero of the Fermi level is set to be the VBM of the host, and the Fermi level range spans the bulk gap.\cite{wang2014microscopic}}
\end{center}
\end{figure}

The large difference in the nature of defects between SnTe and PbTe originates from the different work functions or the distinction of the VBM of these two materials, as the formation energies of charged defects depend explicitly on the Fermi level referenced to the VBM. The VBM of SnTe is 0.5 eV higher than that of PbTe, inducing that SnTe has a lower formation energy of the negatively charged defects than PbTe. Meanwhile, alloying SnTe with Pb can continuously lower the VBM and CBM, so as to modify the conductivity type. So tuning the Pb/Sn ratio in Pb$_{1-x}$Sn$_x$Te alloy could obtain either n-type or p-type conductivity, which is a suitable way to realize bulk insulating and observe the topological surface states.

When doping the magnetism into thin films of TIs, the topological electronic states with a Chern number $\pm 1$ can be formed and the quantum anomalous Hall effect (QAHE) can be realized\cite{yu2010quantized}, which is experimentally observed in Cr doped Bi$_2$Te$_3$ class of TIs\cite{chang2013experimental}. Here in TCIs, Fang {\it et al.}\cite{fang2014large} proposed that when an out-of-plane ferromagnetic order is introduced into a TCI film, the QAHE with a tunable large Chern number can be produced. However, magnetic doping in SnTe-class TCIs is not easily achievable. In the early years, some experimental efforts have been made to study various types of magnetism in SnTe crystals doped with 3d transition metals (TMs), which are found to be degenerate magnetic semiconductor with small magnetic moments for these TM atoms\cite{inoue1981various}. These systems show complex magnetic properties: for example, doping with Mn, Cr and Fe act as ferromagnetic behavior, while doping with Co and Ni are magnetically ineffective\cite{inoue1981various}. More peculiarly, a strong effect of carrier concentration on the magnetic properties of these diluted magnetic semiconductors has been demonstrated, and it can be understood on the basis of the Ruderman-Kittel-Kasuya-Yosida (RKKY) interaction mechanism\cite{story1986carrier}. Recently, some calculations investigated the electronic and magnetic properties of V, Cr or Mn-doped SnTe\cite{liu2013magnetism}. However their doping concentration is very high, and it is likely to damage the TCI phase of SnTe. Another theoretical study comes to a conclusion that the doped TM atoms in SnTe have comparatively high formation energies, and predicts that the uniform TM doping with a higher concentration in SnTe will be difficult unless clustering\cite{wang2016defect}. To date, the effective magnetic doping in SnTe-class TCI materials is still a challenge, and the further investigations of magnetic effect on TCIs are required.

Another important doping is In-doped SnTe, which hosts the superconducting behavior\cite{erickson2009enhanced,sasaki2012odd}. In particular, Sn$_{1-x}$In$_x$Te with an indium content of $x=0.045$ preserving the topological surface states has been proved a topological superconducting state by the signals of surface Andreev bound states in the point-contact spectroscopy, and the critical temperature is 1.2 K\cite{sasaki2012odd,sato2013fermiology}. Furthermore it has recently been reported that the superconducting $T_{\rm C}$ of Sn$_{1-x}$In$_x$Te continues to increase with a higher level of doping, such as 4.5 K with In content of $x=0.45$\cite{novak2013unusual,zhong2013optimizing,polley2016observation}. However, the carrier density does not vary monotonically with Indium content. In fact at high Indium content, Indium has a mixed oxidation state in the system, i.e. it is neither In$^{1+}$ nor In$^{3+}$\cite{haldolaarachchige2015resonance}. So superconducting Sn$_{1-x}$In$_x$Te can not be viewed as a simple hole doped semiconducting material.

\begin{figure*}[bt]
\begin{center}
\includegraphics[width=0.8\textwidth]{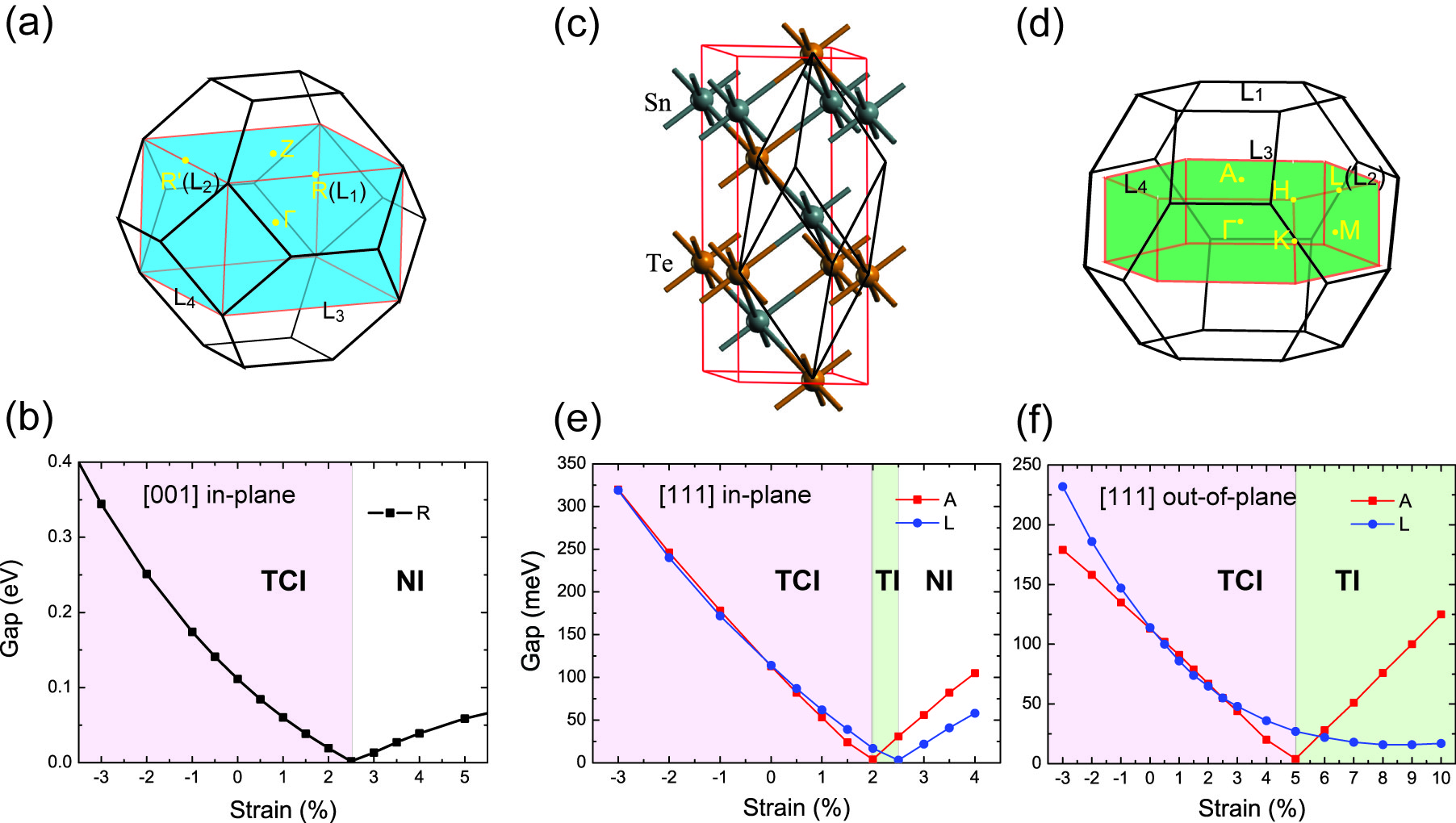}
\caption{\label{fig:fig9} (Color online) (a, b) The effect of [001]-oriented strain on the SnTe bulk. (a) BZs of SnTe for primitive cell (in black) and conventional cell along the [001] direction (in blue). (b) Band gap of bulk SnTe at the $R$ point as a function of in-plane biaxial strain. (c--f) The effect of [111]-oriented strain on the SnTe bulk. (c) Primitive cell (in black) and conventional cell along the [111] direction (in red). (d) BZs of SnTe for primitive cell (in black) and conventional cell along the [111] direction (in green). (e) and (f), Strain-dependent bulk band-gap evolution at the $A$ (red squares) and $L$ (blue circles) points under in-plane and out-of-plane strains, respectively. In (b), (e) and (f), the pink, light green, and white regions indicate the TCI, TI, and NI phases, respectively. Figures (c)--(f) are adapted with permission from Ref.~84, and copyrighted by the American Physical Society.}
\end{center}
\end{figure*}

\section{Strain effect}

Protected by the crystal symmetry and possessing multiple branches of Dirac surface states, TCIs have a much wider range of tunable electronic properties than TIs under various perturbations. Strain is an important controllable method\cite{si2016strain}. Not only can strain drive topological phase transition in the bulk system, but it can also modulate the electronic properties of the Dirac surface states.

\subsection{Strain effect on bulk: topological phase transition}

Strain is an effective way to tune the band gap of semiconductors. For the narrow-gap semiconductors whose VBM and CBM have the opposite parity or bonding types, the strain will influence their levels differently, and even induce the band inversion\cite{liu2014manipulating}. As has been proposed in Bi$_2$Se$_3$ family compounds, strain can induce the topological phase transition from a normal insulator (NI) to a TI and vice verse\cite{young2011theoretical,liu2011anisotropic,liu2014manipulating}. This manipulation can also be applied to TCIs. As mentioned by Hsieh {\it et al.}\cite{hsieh2012topological}, with the lattice constant decreasing, PbTe can be changed from a NI to a TCI. Such topological phase transition also occurs for other lead chalcogenides under external pressure or volume compression\cite{barone2013pressure,barone2013strain}; while for the volume expansion, SnS and SnSe can be transformed from an ambient pressure TCI to a topologically trivial insulator\cite{sun2013rocksalt}.

As has been known to us, there are four $L$ points in the bulk BZ; the isotropic strain can not distinguish them and will change the band orders of four valleys simultaneously. So the isotropic strain, such as hydrostatic pressure for the rocksalt structure, can only induce the topological phase transition between a NI and a TCI. However, an uniaxial or biaxial strain, which could be applied in the epitaxial films, can offer a diverse manipulation of the electronic properties. A systematic study of SnTe with various in-plane biaxial strains perpendicular to the [001] direction has been made by Qian {\it et al.}\cite{qian2014topological} Since the unit cell fitting for this type of strain is along the [001] direction, there is a folding of bulk BZ. As depicted in Fig. 9(a), four $L$ points in the original BZ (polyhedron encircled by black lines) are folded to $R$ and $R^{\prime}$ in the new BZ (blue tetragonal cell). The in-plane biaxial strain does not break the $C_4$ rotation, so $R$ and $R^{\prime}$ are still equivalent. The calculated band gap at $R$ as a function of biaxial strain is shown in Fig. 9(b). As the strain increases, the energy gap initially decreases, and then opens up again. Above 2.5$\%$ strain, band inversion disappears, and the system is transformed from a TCI to a NI.

A special strain which can distinguish the four $L$ points is the [111]-oriented strain. Zhao {\it et al.}\cite{zhao2015tuning} studied the band evolution of SnTe with the in-plane biaxial strain and out-of-plane uniaxial strain in the [111] direction. The unit cell and BZ are shown in Figs. 9(c) and (d). Due to the enlarged unit cell, the BZ shrinks from the face-centered-cubic case to a hexagonal prism. Consequently, the $L_1$ point in the original BZ folds to the $A$ point in the new BZ, and the band at $A$ will change differently from those at other three $L$ points under strain. The band-gap evolutions at the $A$ and $L$ points under the two types of strains (the in-plane biaxial strain and the out-of-plane uniaxial strain) are shown in Figs. 9(e) and (f), respectively. As the in-plane strain increases, though the band gaps at both $A$ and $L$ reduce to zero then reopen again, the gap closing will first occur at the $A$ point. So except for TCI and NI phases, there is a TI phase for the in-plane strain level between +2$\%$ and +2.5$\%$. For the out-of-plane strain, the band-gap closing and reopening only occur at the A point, so below +10$\%$, there are two topological phase: TCI and TI, where the TI phase could appear in the range of $+5 \% < \varepsilon_{\perp} < +10 \% $.

\subsection{Strain effect on (001) surface}

Here, maintaining the TCI phase, we discuss the effect of strain on the (001) surface states. The Dirac points on the (001) surface of SnTe are not pinned to TRIMs. A mechanical strain can shift the Dirac point positions in the momentum space just like an electromagnetic field acts on an electron\cite{tang2014strain}. Qian {\it et al.}\cite{qian2014topological} studied the strain-dependent electronic properties of SnTe (001) nanomembranes. Here we take the band structure of a 51-layer (001) membrane under the in-plane biaxial strain as an example. As shown in Fig. 10, the compressive strain drives the Dirac points of surface states to move away from the $\bar{X}$ points, while the tensile strain makes them close to the $\bar{X}$ points. In addition, the strain will influence the penetration length of the surface states like the case of TIs. Qian {\it et al.}\cite{qian2014topological} found that the compressive (tensile) strain can reduce (increase) the surface state penetration length, so the tensile strain will cause a hybridization gap for the SnTe thin film, such as the surface band for the 51-layer membrane under the biaxial strain of +1$\%$ [right panel in Fig. 10]. Such shift of Dirac point position in k space under strain has been realized in experiments\cite{zeljkovic2015strain}, which establishes a tunable platform for the strain-based applications.

\begin{figure}[ht]
\begin{center}
\includegraphics[width=0.47\textwidth]{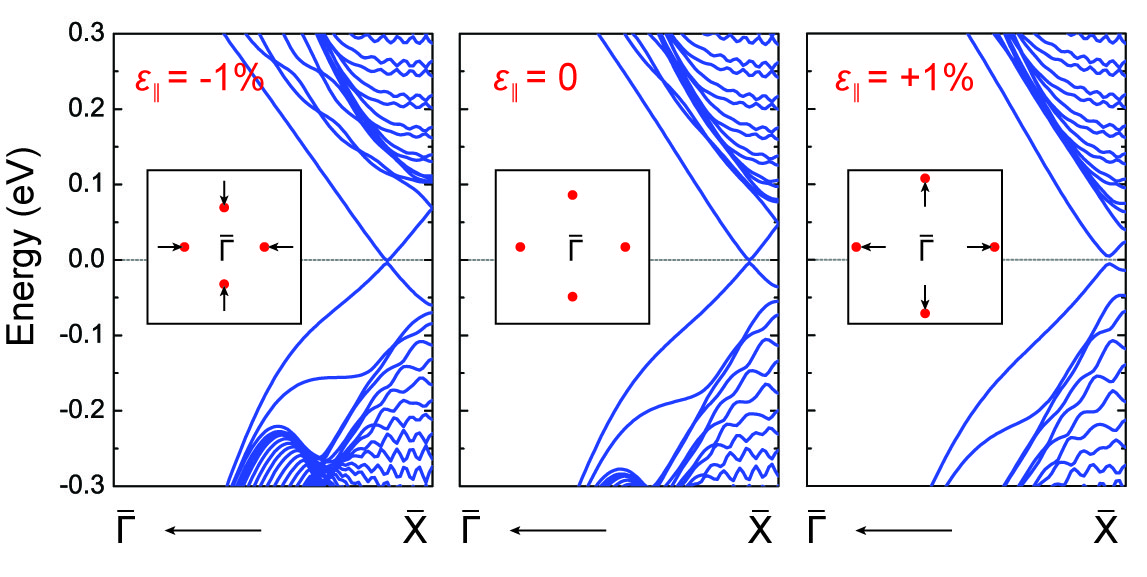}
\caption{\label{fig:fig10} (Color online) First-principles band structures of a 51-layer SnTe (001) nanomembrane under biaxial strain of -1$\%$, 0 and +1$\%$. Insets schematically indicate the shift of Dirac points in the surface BZ under strain.}
\end{center}
\end{figure}

The biaxial strain will equally shift the four Dirac points of the (001) surface states. However, a structural distortion with a relative displacement of atoms along the [110] direction on the (001) surface can break one of two mirror planes, and generate a mass to the Dirac cone. As a result, the two Dirac points along [$1\bar{1}0$] direction are gapped, but the other two along [110] remain gapless, which has been demonstrated by the STM measurements of the coexistence of zero-mass Dirac fermions with massive Dirac fermions\cite{okada2013observation}.

\subsection{Strain effect on (111) surface}

The Dirac surface states on the (111) surface are noninteracting and located at well-separated TRIMs (i.e., $\bar{\Gamma}$ and $\bar{M}$) in the surface BZ. Most importantly, differing from the valleys on the (001) surface, the $\bar{\Gamma}$ and $\bar{M}$ valleys appear highly distinct for crystal symmetry, and thus their band topology is sensitively differentiated by mechanical deformation as mentioned above. In addition, strain effect could also alter the valley-based landscape in the (111) surface.

\begin{figure*}[ht]
\begin{center}
\includegraphics[width=0.8\textwidth]{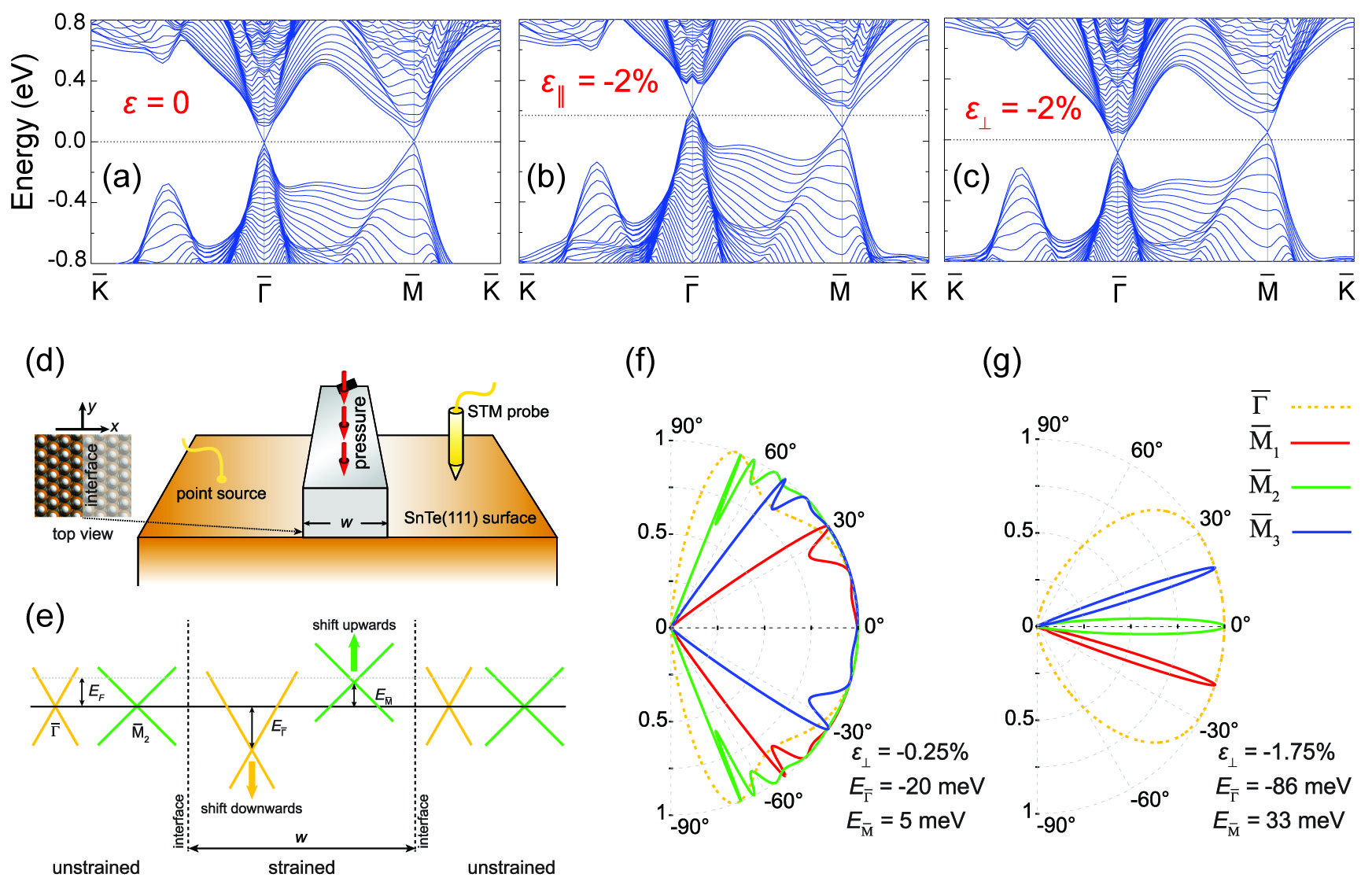}
\caption{\label{fig:fig11} (Color online) (a--c) DFT-calculated surface band structures of SnTe (111) surface under no strain, in-plane biaxial strain of -2$\%$ and out-of-plane uniaxial strain of -2$\%$, respectively. The horizontal dotted lines denote the Fermi levels under different strains, and all bands have set the Fermi level in the unstrained system as the zero reference for energy. (d) Schematic of a strain-induced nanostructure in the SnTe (111) surface grown on a substrate. (e) Schematic of the energy spectra in the strain-induced nanostructure shown in (d). (f) and (g), Transmission probabilities $T(\theta)$ of massless Dirac fermions at the $\bar{\Gamma}$ and $\bar{M}$ valleys at the strain levels of $\varepsilon_{\perp}=-0.25{\%}$ and $\varepsilon_{\perp}=-1.75{\%}$, respectively.\cite{zhao2015tuning}}
\end{center}
\end{figure*}

Maintaining the TCI phase of the entire system, Zhao {\it et al.}\cite{zhao2015tuning} studied the strain-dependent Dirac valley evolution on the SnTe (111) surface under different strains. Figs. 11(a--c) show the DFT-calculated surface band structures under no strain, in-plane biaxial strain of -2$\%$ and out-of-plane uniaxial strain of -2$\%$, respectively. In the absence of strain [Fig. 11(a)], the $\bar{\Gamma}$ and $\bar{M}$ valleys are degenerate in energy; while the compressive strain drives the surface bands at $\bar{\Gamma}$ and $\bar{M}$ to shift oppositely with reference to the Fermi level [Figs. 11(b) and (c)]. In particular, for the out-of-plane compressive strain, the opposite shift of the surface bands at $\bar{\Gamma}$ and $\bar{M}$ also applies with reference to the unstrained Fermi level. Such properties can be used to design the strain-engineered nanodevices for dynamic valley control. As depicted in Fig. 11(d), a local compressive strain in the SnTe (111) surface can be generated by applying perpendicular pressure using a piezoelectric-material-based actuator. Under out-of-plane compressive strain, the Dirac cone at ${\bar{\Gamma}}$ shifts downwards, while the cones at ${\bar{M}}$ shift upwards, as schematically illustrated in Fig. 11(e). Therefore, with increasing strain, the Fermi level in the strained region will approach the Dirac point at $\bar{M}$, then goes across the valence band. The opposite shift in energy of Dirac cones at ${\bar{\Gamma}}$ and $\bar{M}$ in the strained region will lead to different transport behaviors for the massless Dirac fermions at these four valleys. Note that the strain-driven shift of Dirac point in energy also appears in Bi$_2$Se$_3$-class TIs\cite{zhao2012design}, but there is only one Dirac cone located at $\bar{\Gamma}$ for them.

The transmission probabilities $T(\theta)$ through the strain-induced nanostructure with two different strain levels are shown in Figs. 11(f) and (g). It can be seen that for the $\bar{\Gamma}$ valley, high transmission probabilities can always be maintained over a large angular range. This is because the Dirac point shifts downward and the scattering events always stay in the intraband tunneling regime. On the contrary, the Dirac points at the $\bar{M}$ valley shift upward with increasing strain, and the Fermi level in the strained region will gradually approach the Dirac point of zero carrier concentration. Eventually, the scattering of Dirac fermions at the $\bar{M}$ valley will be driven into the interband tunneling regime. Particularly in Fig. 11(g), when the Fermi level exactly crosses the Dirac points at $\bar{M}$ in the strained region, strong beam collimation along three different directions (i.e., $\theta=0$, $\pm 18.2^{\circ}$) can be clearly seen. Deviating from these specific directions, it is possible to achieve the pure current of the $\bar{\Gamma}$ valley. So the strong valley-selective filtering for massless Dirac fermions can be accomplished by dynamically applying local external pressure, which pave the way for strain-engineered nanoelectronic and valleytronic applications in TCIs.

\begin{figure*}[ht]
\begin{center}
\includegraphics[width=0.8\textwidth]{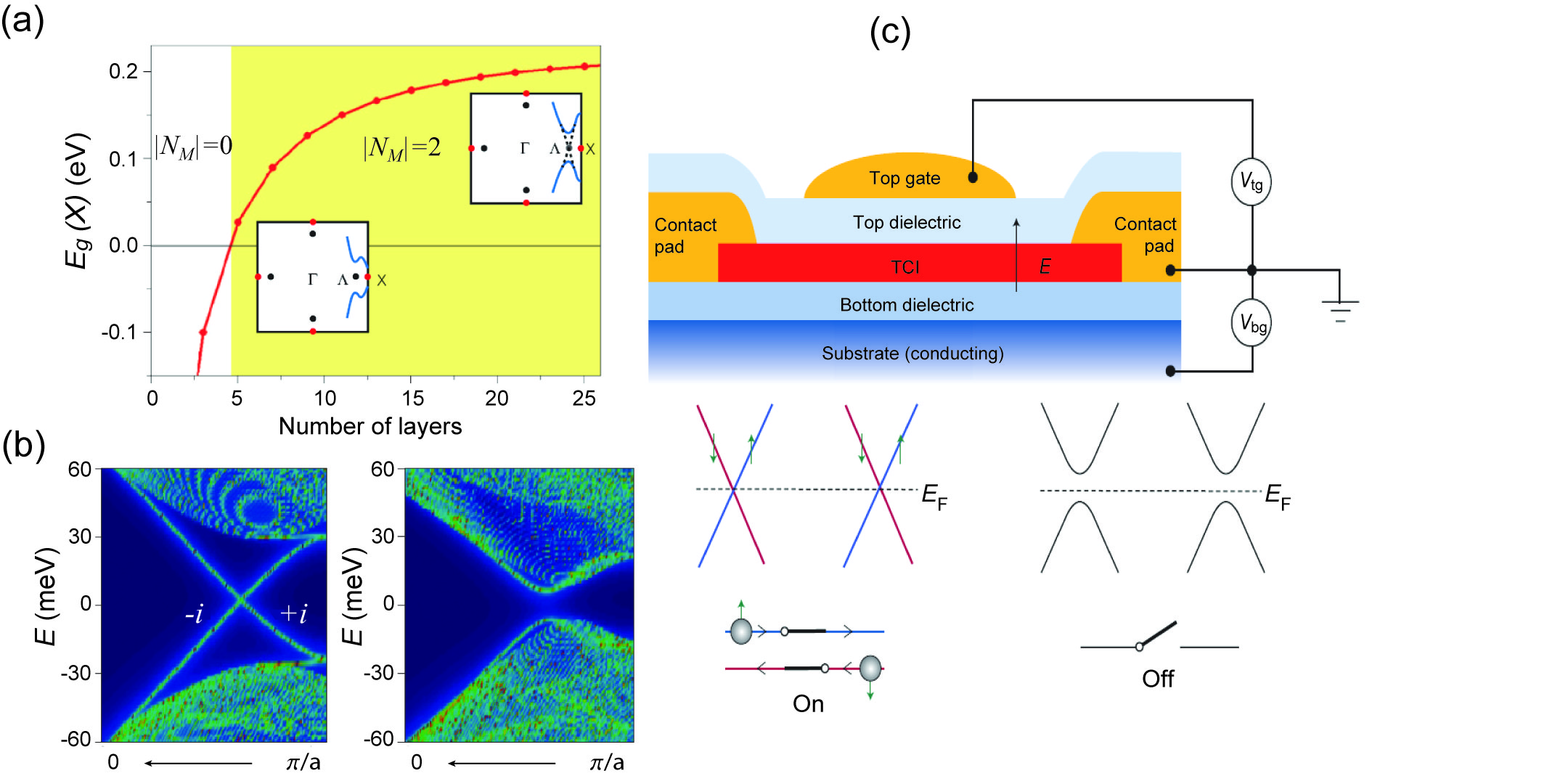}
\caption{\label{fig:fig12} (Color online) (a) Band gap of SnTe at the $X$ point as a function of film thickness. The yellow region indicates a 2D TCI phase. (b) Edge states with mirror eigenvalues of an 11-layer SnTe film (left) and electric-field-induced gap (right). (c) Proposed topological transistor device for using an electric field to tune charge and spin transport. Without the electric field, the spin-filtered edge states are gapless and it is in the ``on'' state (left). Applying a perpendicular electric field, the edge states are gapped and it is in the ``off'' state (right).\cite{liu2014spin}}
\end{center}
\end{figure*}

\section{Thickness engineering}

When the thickness of SnTe film decreases, the top and bottom surface states will hybridize to open up an energy gap at the Dirac points. However, the hybridization is complex and multiple novel properties can emerge.

\subsection{2D TCI in (001) thin films}

The Dirac point of SnTe (001) surface states is deviated from the $\bar{X}$ point in the surface BZ; while at $\bar{X}$, there is an inverted band gap, i.e., the valence band (conduction band) is derived from the cation (anion). When the thickness of SnTe (001) film is below the penetration length of surface states, the top and bottom surface states will hybridize and result in an energy splitting between the bonding and anti-bonding states. The conduction and valence bands of the 2D film at $X$ will come from the bonding state of the anion at energy $E_{\rm A+}(X)$ and anti-bonding state of the cation at energy $E_{\rm C-}(X)$\cite{liu2014spin}. However, the band ordering of $E_{\rm A+}(X)$ and $E_{\rm C-}(X)$ depends on the competition between the hybridization of the two surfaces and the inverted gap of each surface. For thick films, the hybridization is weak so that $E_{\rm A+}(X) > E_{\rm C-}(X)$, which inherits the inverted band gap of the 3D limit. While the strong hybridization in the thin film drives $E_{\rm C-}(X)$ higher than $E_{\rm A+}(X)$, leading to a trivial phase. Considering the SnTe (001) films with an odd number of atomic layers, which are symmetric under the reflection $z \rightarrow -z$ about the (001) plane in the middle, the band gap [$E_{\rm A+}(X)-E_{\rm C-}(X)$] at the $X$ point as a function of film thickness is shown in Fig. 12(a)\cite{liu2014spin}. Above five layers, the gap of SnTe at $X$ is inverted and increases with the thickness. Due to the (001) mirror symmetry of the film, one can define the mirror Chern number of 2D film as in the case of 3D TCI. Liu {\it et al.}\cite{liu2014spin} found that the thick (001) film with an inverted gap at $X$ is a 2D TCI with mirror Chern number $|N_M|=2$ [Fig. 12(a)].

An important result of 2D TCI with mirror Chern number $|N_M|=2$ is that there are two pairs of counter-propagating edge states in the band gap. As shown in the left panel of Fig. 12(b), in a strip structure of 11-layer SnTe thin film, the edge states with opposite mirror eigenvalues cross each other at the edge of BZ, but are not located at TRIM. Here the Dirac-like edge states are protected by the (001) mirror symmetry, rather than the time-reversal symmetry as in the quantum spin Hall insulator. So applying a perpendicular electric field, which breaks the mirror symmetry, will generate a band gap in these edge states, as is shown in the right panel of Fig. 12(b). This unique property that the conductance of edge states is easily and widely tunable by an electric-field-induced gap instead of carrier depletion can be utilized to propose a topological transistor device made of dual-gated TCI thin films, as depicted in Fig. 12(c). With high on/off speed, the device can control the coupled charge and spin transport by purely electrical means.\cite{liu2014spin}

A similar mechanism appears in PbTe/SnTe superlattices along the [001] direction\cite{yang2014weak}. The folding of BZ due to the superlattice structure projects two $L$ points in the original BZ to a single point $L^{\prime}$, and their hybridizations form the bonding and anti-bonding states. The coupling strength between the states at equivalent $L$ points depends on the relative thickness or the ratio $m/2n$ in (PbTe)$_m$(SnTe)$_{2n-m}$ superlattice. For a large range of this ratio, the strong coupling induces a band inversion between the bonding state of conduction band and anti-bonding state of valence band. Thanks to reduction in the number of $L$ points from four to two, the band inversion at two $L^{\prime}$ will lead to a weak TI phase in PbTe/SnTe superlattices\cite{yang2014weak}.

\subsection{Quantum spin Hall states in (111) thin films}

Another important film is the (111) thin film, which has been grown epitaxially in recent experiments\cite{yan2014experimental,polley2014observation,wang2015molecular}. Different from (001) surface states, four Dirac cones of (111) surface states are centered at TRIMs: one is at the $\Gamma$ point and the other three are at the $M$ points, and they are not all equivalent. When the thickness of (111) film decreases, the hybridization strength between the top and bottom surface states at $\Gamma$ and $M$ will be different. In fact, the penetration length of the surface states at $M$ is much larger than that at $\Gamma$, hence the hybridization-induced gap at $M$ is correspondingly larger than the one at $\Gamma$ by orders of magnitude\cite{liu2015electrically}. As a result of this inequivalence between $\Gamma$ and $M$, the properties of SnTe (111) films are mainly determined by the low-energy physics at the $\Gamma$ point. The magnitude and sign of the gap at $\Gamma$ will depend on the film thickness, thus leading to a topological phase transition as a function of the film thickness\cite{liu2015electrically}. Taking the films with an even number of layers as an example [Fig. 13(a)], when the thickness decreases, both the fundamental gap $E_g$ and the gap at the $\Gamma$ point $E_g(\Gamma)$ exhibit a oscillatory behavior, indicating a sign change of the Dirac mass and the $Z_2$ topological phase transition. As shown in the inset of Fig. 13(a), the conduction bands of the 12-layer thin film are mainly derived from Sn orbitals, while they are mainly from Te orbitals for the 14-layer thin film, suggesting that 14-layer thin film is a quantum spin Hall (QSH) state. This result can be confirmed by the edge state calculations. As shown in Fig. 13(b), for a 14-layer thin film, the linear-dispersion gapless states exist in the bulk gap with a Dirac point at $\bar{\Gamma}$, while there are no such edge states in the 12-layer thin film. From Fig. 13(a), all the thin films with an even number of layers between 14 and 30 layers are QSH insulators.\cite{liu2015electrically}

\begin{figure}[ht]
\begin{center}
\includegraphics[width=0.48\textwidth]{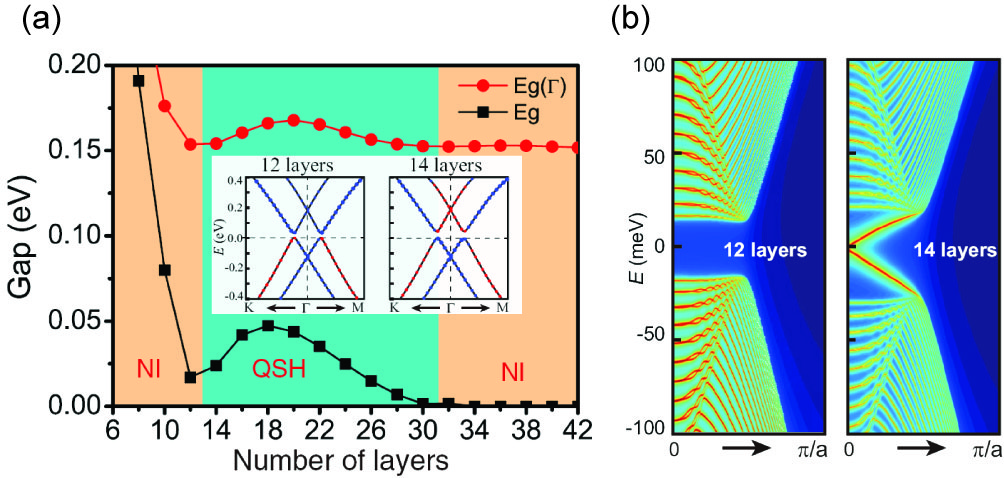}
\caption{\label{fig:fig13} (Color online) (a) The fundamental gap $E_g$ and the gap at the $\Gamma$ point $E_g(\Gamma)$ as functions of the film thickness with even number layers. The blue region indicates a quantum spin Hall (QSH) state, and the orange region represents a normal insulator (NI). The red (blue) dots in the inset denote the contribution of Te (Sn) atoms. (b) The edge states of a 12-layer thin film (left) and 14-layer thin film (right).\cite{liu2015electrically}}
\end{center}
\end{figure}

The similar results apply to the odd-layer thin film\cite{liu2015electrically}, and even the SnSe (111) film\cite{safaei2015quantum}. The mechanism for realizing the QSH phase in TCI thin films through the intersurface coupling is similar to the thin films of Bi$_2$Se$_3$-class TIs\cite{linder2009anomalous,liu2010oscillatory,lu2010massive}. However, due to the much larger penetration length of the SnTe/SnSe surface states, the hybridization gap is much larger than that in Bi$_2$Se$_3$ class of TIs.

\subsection{Electric field and helicity control}

The Dirac fermion of the surface states in the topological materials has the property of spin-momentum locking, which can also be defined as the helicity degree of freedom. As depicted in Fig. 14(a), in the film of topological materials, two surface Dirac cones located on the top and bottom surfaces are mirror images of each other obeying inversion symmetry and have opposite helicity. Specifically, the top surface has the left-handed (LH) electron and right-handed (RH) hole states, while the bottom surface has the RH electron and LH hole states. For the thinner film [Fig. 14(b)], the two Dirac cones interact with each other and open a hybridization gap. Meanwhile, the twofold helicity-degenerate band structures sustain both LH and RH electron (or hole) states of Dirac fermions. However, the helicity degeneracy can be removed in a controllable way by breaking the spatial inversion symmetry of the film, e.g., by applying an external electric field [Fig. 14(c)]. This feature allows the control over the helicity degree of freedom of Dirac fermions.

\begin{figure}[ht]
\begin{center}
\includegraphics[width=0.47\textwidth]{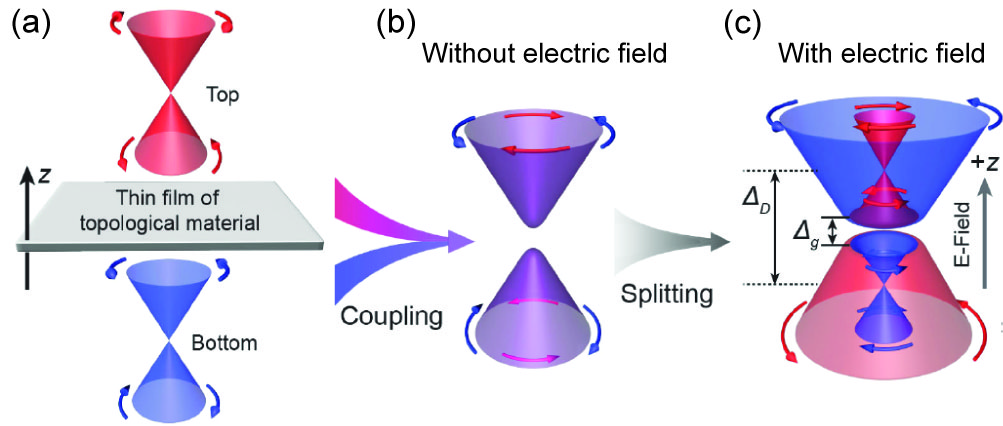}
\caption{\label{fig:fig14} (Color online) A simplified model of electrically tuned band evolution in a thin film of topological material, where the helicity degree of freedom in the system can be controlled by a perpendicular electric field.\cite{zhao2015electronic}}
\end{center}
\end{figure}

Taking an odd-layer SnTe (111) film as examples, Zhao {\it et al.}\cite{zhao2015electronic} studied the helicity-resolved band structures and transports. They found that a moderate electric field can induce a giant helicity splitting to the degenerate bands in the vicinity of the $\bar{\Gamma}$ point. To explore helicity-dependent manipulation of Dirac fermions, a dual-gated heterostructure is exploited in SnTe (111) film, as shown in Fig. 15(a). For Fabry-P\'{e}rot-type model\cite{zhao2015electronic}, when tuning the Fermi level in the dual-gated region to cross the upper Dirac point, a strong direction-dependent helicity-selective transmission is indicated in Fig. 15(b). The RH electron wave has very high transmission probability over a large angular range, whereas a sharp transmission peak appears at normal incidence for the LH electron wave. Such pronounced helicity-selective transmission stems from the fact that the intrahelical scattering dominates over the interhelical one in the transmission events. When changing the width of dual-gated region and tuning the Fermi level, more helicity-resolved functionalities, including helical negative refraction, birefraction, and electronic focusing, are demonstrated\cite{zhao2015electronic}. Like the spin manipulation in thin film of magnetically doped TIs\cite{zhao2013field}, here the electrically tunable helicity-resolved control of Dirac fermions could open up a promising field for helicity-based electronics.

\begin{figure}[ht]
\begin{center}
\includegraphics[width=0.47\textwidth]{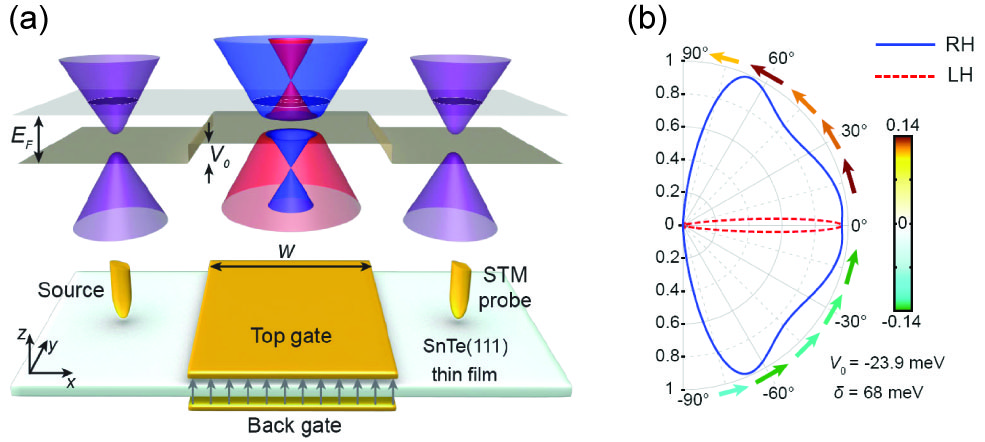}
\caption{\label{fig:fig15} (Color online) (a) Schematic design of a dual-gated heterostructure tuned by the electric field. The incident helicity-degenerate Dirac electrons emitted from the source, and the transmitted helical electron waves are detected by the STM probe. (b) Pronounced helicity filtering when the Fermi level crosses the upper Dirac point. The colorful arrows illustrate the spin orientations at different transmission angles, where the color scale gives the out-of-plane spin component.\cite{zhao2015electronic}}
\end{center}
\end{figure}

\section{Conclusions and outlook}

In this article, we have reviewed the electronic properties of SnTe-class TCI materials, from the intrinsic topology, orientation-dependent surface states, to tunable properties under various perturbations. Protected by the mirror symmetry and characterized by the mirror Chern number, SnTe-class TCI materials have multiple branches of Dirac surface states, whose types depend on the surface orientations and even the surface reconstructions. Chemical doping can realize the control of carrier types or concentrations, the topological superconductivity, and the large-Chern-number QAHE. Strain is an effective way to tune the electronic properties of IV-VI materials, including driving topological phase transition for the bulk system, shifting the Dirac point position of the surface states in the momentum space or energy space. The latter could be used for the valleytronic applications. For the thin film of SnTe-class materials, the thickness of film can influence the hybridization strength, and create diverse topological phases, for instance, 2D TCI, weak TI, QSH states, and NI. Among them, the (001) thin film with 2D TCI phase could be utilized to design a topological transistor device with a high on/off speed, and more generally, the helicity degree of freedom of Dirac fermions in the thin film can be controlled by an external electric field to realize the helicity-resolved functionalities.

These novel properties and potential applications are not limited to the IV-VI materials. In fact, more TCIs are proposed theoretically\cite{kargarian2013topological,hsieh2014topological,liu2014topological,jadaun2013topological,slager2013space,dong2015classification}, and a great deal of studies are needed to search for new TCI materials and explore their applications. Moreover, the crystal symmetry protection has been extended from the insulators to semimetals, such as the Dirac semimetal\cite{young2012dirac,wang2012dirac,liu2014discovery,wang2013three,liu2014stable,weng2016topological} and nodal-line semimetal\cite{kim2015dirac,burkov2011topological,bian2015topological,bian2015drumhead,weng2016topological}, which possess more exotic effects\cite{hosur2013recent,bian2015topological,bian2015drumhead}. The interplay between the electronic topology and crystal symmetry has broadened the field of topological phase, and paves a path for developing the quantum electronic and spintronic devices.





\providecommand{\noopsort}[1]{}\providecommand{\singleletter}[1]{#1}%

\end{document}